\documentclass[fontsize=11pt,twoside=semi,numbers=endperiod]{scrartcl}
\KOMAoptions{headinclude=true,footinclude=false,index=toc}
\usepackage[paperwidth=150mm,paperheight=179.35mm,width=110mm,totalheight=165.35mm,includehead,top=4mm,inner=20mm]{geometry} 

\usepackage{amsmath}
\usepackage{amssymb}

\usepackage{lmodern} 

\usepackage[ansinew]{inputenc}
\usepackage[T1]{fontenc}
\usepackage[english,ngerman]{babel}
\usepackage{textcomp} 
\usepackage[pdftex]{color,graphicx} %
\usepackage[rel]{overpic} 
\usepackage{bbm}
\usepackage{cite} 
\usepackage{ellipsis} 
\usepackage{microtype} 
\usepackage{simplewick} 
\usepackage{scrpage2}
\usepackage[%
	pdftex,
	colorlinks,citecolor=blue,filecolor=blue,linkcolor=blue,urlcolor=blue,%
	bookmarks=true,%
	bookmarksopen=false,%
	bookmarksnumbered=true,%
	pdfpagemode=UseNone,
    pdfstartview={XYZ null null 1.00},
	pdffitwindow=true,
    pdfview={XYZ null null null},
    pdfpagelayout=SinglePage,
    pdfdisplaydoctitle=true, 
    plainpages=false,
    pdfpagelabels=true,
	pdfkeywords={vacuum energy, zero-point oscillation, zero-point energy, canonical quantization, substrate-less field, cosmological constant problem, electro-weak phase transition, Casimir effect},
	pdftitle={Canonical quantization of substrate-less fields},
    pdfauthor={Astrophysical Institute Neunhof}]
	{hyperref}

\setkomafont{disposition}{\rmfamily} 
\addtokomafont{section}{\large\bfseries} 
\addtokomafont{subsection}{\normalsize\bfseries} 
\addtokomafont{pagenumber}{\LARGE}
\addtokomafont{caption}{\small}

\deffootnote[1em]{1em}{1em}{\textsuperscript{\thefootnotemark}\ }

\newcommand{\ggitem}[2][$\ast $]{\makebox[0em][r]{#1\;}{#2}\newline }%
\newenvironment{ggitemize}{\vspace{0.2\baselineskip} \begin{addmargin}[1.2em]{0em}}{\vspace{-0.8\baselineskip} \end{addmargin}}

\newcommand{\ggie}{\mbox{i.\,e.}\ } 
\newcommand{\ggeg}{\mbox{e.\,g.}\ } 
 
\newcommand{\bz}[1]{\nolinebreak\hspace{0em}\nolinebreak{}#1\hspace{0em}}
\newcommand{\al}{\iflanguage{english}{``\nolinebreak\hspace{0em}}{\glqq\nolinebreak\hspace{0.1em}}\nolinebreak}
\newcommand{\ar}{\nolinebreak\iflanguage{english}{\hspace{0em}\nolinebreak{}''\ }{\hspace{-0.45em}\nolinebreak\grqq\thickspace}}
\newcommand{\arp}{\nolinebreak\iflanguage{english}{\hspace{0em}\nolinebreak{}''}{\grqq}}

\newcommand{\ggstackrel}[2][0]{\hspace{0.36em}\hspace{#1em}&=\hspace{-#1em}\hspace{-1.16em}\stackrel{#2}{\phantom{=}}} 
\newcommand{\textstackrel}[2][=]{\raisebox{-1mm}[0mm][0mm]{$\stackrel{#2}{#1}$}} 

\newcommand{\inte }{\int \limits }

\newcommand{\mklammerp}{\mbox{\hspace{.4em}\raisebox{1.1mm}[0mm][0mm]{$-$}\hspace{-1em}\raisebox{-.4mm}[0mm][0mm]{\scalebox{.8}[.55]{$($}\scalebox{.8}{$+$}\scalebox{.8}[.55]{$)$}}\hspace{.2em}}}
\newcommand{\pklammerm}{\mbox{\hspace{.4em}\raisebox{1.1mm}[0mm][0mm]{\scalebox{.8}{$+$}}\hspace{-1em}\raisebox{-.4mm}[0mm][0mm]{\scalebox{.8}[.55]{$($}\raisebox{-.4mm}[0mm][0mm]{$-$}\scalebox{.8}[.55]{)}}\hspace{.2em}}}

\DeclareMathOperator{\dif }{d}

\clearscrheadfoot 
\rohead[\vspace{-2mm}\par\pagemark ]{\vspace{-2mm}\par\pagemark}
\lehead[\vspace{-2mm}\par\pagemark ]{\vspace{-2mm}\par\pagemark}
\rehead{ }
\lohead{ }

\begin{document}
\selectlanguage{english}
\thispagestyle{empty}
\raggedbottom  
\vspace*{3mm}
\centerline{\bfseries\Large Canonical quantization of}
\vspace*{.5mm}
\centerline{\bfseries\Large elementary fields}
\vspace*{2mm}
\centerline{{\bfseries Gerold Gr\"{u}ndler}\,\footnote{\href{mailto:gerold.gruendler@astrophys-neunhof.de}{email:\ gerold.gruendler@astrophys-neunhof.de}}}
\vspace*{1mm}
\centerline{\small Astrophysical Institute Neunhof, N\"{u}rnberg, Germany} 
\vspace*{3mm}
\noindent\parbox{\textwidth}{\small An inconsistency of quantum field theory, regarding the signs of vacuum energy and vacuum pressure of elementary fields versus non\bz{-}elementary fields (like \ggeg  phonon fields), is pointed out. An improved law for the canonical quantization of fields is presented, which is based on the distinction between elementary and non\bz{-}elementary fields. Remarkably, the improved quantization method removes not only the inconsistency of quantum field theory, but at the same time solves the (old) cosmological constant problem for all fields of the standard model of elementary particles (but not for the hypothetical inflaton fields), without compromising any of the achievements of established quantum field theory.} 
\vspace*{1mm}

\noindent{}{\small PACS: 03.70.+k, 98.80.Es} 
\vspace*{1mm}

\noindent{}{\small Keywords: vacuum energy, zero\bz{-}point oscillation, zero\bz{-}point energy, canonical quantization, substrate\bz{-}less field, cosmological constant problem, electro\bz{-}weak phase transition, Casimir effect} 

\section{Overview} 
Elementary fields (like the electromagnetic field, or the electron\bz{/}positron field, or the gluon fields, or\,\dots\,) have infinitely many degrees of freedom, while non\bz{-}elementary fields (like phonon fields, which will be our standard example for non\bz{-}elementary fields) have only a finite number of degrees of freedom (the $3N-6$ normal vibrational modes in case of the phonon field of an $N$\bz{-}atom solid). Due to this difference, the zero\bz{-}point values of the energydensity\bz{-}stress tensors of quantized elementary fields diverge and need regularization and renormalization, while the zero\bz{-}point values of the energydensity\bz{-}stress tensors of quantized non\bz{-}elementary fields are always finite. 

This difference results into an inconsistency within quantum field theory (QFT), regarding the $\pm $ signs of vacuum energy and vacuum pressure, which is pointed out in section\;\ref{Absch:inconsistQFT}\,. The inconsistency can be removed by a slight but important modification of QFT: A re\bz{-}definition of the rule of canonical field quantization. 

In conventional QFT, there are two different rules for canonical quantization: One for boson fields, and one for fermion fields. The mentioned inconsistency vanishes, if the quantization rule is furthermore split into rules for elementary fields and rules for non\bz{-}elementary fields, such that canonical quantization in total is differentiating between four types of fields: Elementary boson fields, non\bz{-}elementary boson fields, elementary fermion fields, non\bz{-}elementary fermion fields. 

The improved law of canonical quantization of fields is stated in section\;\ref{Absch:imprcanquant}\,. The improved law essentially has the consequence that the zero\bz{-}point oscillations, and hence the zero\bz{-}point energies and zero\bz{-}point pressures, of elementary quantum fields vanish. This solves at the same time the (old) cosmological constant problem, \ggie  the enormous mismatch between the vacuum energy expected according to QFT, and the astronomically observed vacuum energy. 

As the Higgs field --- different from all other fields of the standard model of elementary particles --- is not linear, the effect of the improved quantization law on this field is not immediately obvious. Therefore the Higgs field is separately discussed in section\;\ref{sec:higgs}. The improved quantization law comes with a trade\bz{-}off: If the hypothetical inflaton fields should really exist and really have triggered one or several cosmic inflations during the evolution of our universe, then they must be basically different from the fields of the standard model of elementary particles, and obey different laws of nature. This complication is briefly discussed at the end of section\;\ref{sec:higgs}\,. 

Even if the zero\bz{-}point oscillations of elementary fields are eliminated, shouldn't the cosmological constant problem turn up again in each order of perturbative treatment of interacting fields? In section\;\ref{Absch:vacfluc} it will be explained, why this is not the case. 

As the Casimir effect is often cited as evidence for the reality of zero\bz{-}point oscillations of elementary fields, section\;\ref{sec:caseff} is dedicated to the refutation of that argument.

Eventually the findings of this article will be summarized in section\;\ref{sec:conclus}\,. 

Upfront in section\;\ref{absch:canquant} conventional canonical field quantization and the cosmological constant problem are briefly delineated, to (a) introduce our notations and (b) prepare the argumentation of sections \ref{Absch:inconsistQFT} and \ref{Absch:imprcanquant}\,. 

\section{Conventional canonical quantization, and the (old) cosmological constant problem}\label{absch:canquant}
Heisenberg\!\cite{Heisenberg:umdeutung} found the rules of point\bz{-}particle quantum mechanics by ingenious guessing. When Dirac\!\cite{Dirac:canquant} analyzed Heisenberg's strange non\bz{-}commutative matrix mechanics, he detected this general correlation between the observable quantities of classical physics, and the non\bz{-}commutative algebra of their quantum\bz{-}mechanical counterparts: 
\begin{subequations}\label{msmngnsd}\begin{align}
A_sB_t-B_tA_s&\! =\! -i\hbar\big\{ A_s^\text{classical}\; ,\: B_t^\text{classical}\big\}\raisebox{-1.5ex}{\scriptsize Poisson\,brackets}\\ 
&\! =\! -i\hbar\delta _{st}\quad\mbox{\parbox[t]{.7\linewidth}{if $A_s^\text{classical}$ is the canonical con-\\ jugate momentum of $B_s^\text{classical}$}}\hspace*{-5em}
\end{align}\end{subequations} 

The quantities on the left\bz{-}hand side are non\bz{-}commutative operators (q\bz{-}numbers), the quantities on the right\bz{-}hand side are commutative classical quantities (c\bz{-}numbers). 

Dirac's method of canonical quantization was generalized to boson\!\cite{Born:qm2,Dirac:quantemfeld} and fermion\!\cite{Jordan:fermfeldquant} fields $\phi (x)$ and their canonical conjugate momentum densities 
\begin{align}\label{wehgfsqaa} 
\pi _a(x)\equiv\frac{\partial\mathcal{L}}{c\partial\big(\partial _0{\phi ^a}(x)\big)} 
\end{align} 
due to the equal\bz{-}time 
\begin{align}
&\text{canonical quantization rule for boson (fermion) fields:}\notag\\ 
&\pi _a(t,\boldsymbol{x})\phi ^b(t,\boldsymbol{y})\mklammerp\phi ^b(t,\boldsymbol{y})\pi _a(t,\boldsymbol{x})=-i\hbar\delta _{ab}\,\delta ^{(3)}\! (\boldsymbol{x}-\boldsymbol{y})\notag\\ 
&\text{All other (anti)commutators are zero.}\label{ksmgnsfngf}
\end{align} 
Here $\mathcal{L}$ is the field's Lagrangian, the indices $_a$ and $_b$ are indicating space\bz{-}time or spinor components, and $x=(t,\boldsymbol{x})$ is a space\bz{-}time point. While the quantization rule for boson fields is nothing but a straightforward generalization of Dirac's quantization rule \eqref{msmngnsd}, the plus sign for fermion fields was a basically new feature. Pauli proved in his spin\bz{-}statistic theorem\!\cite{Pauli:Spinstattheor} that the respective plus and minus signs are indispensable, if the field theories of bosons and fermions both shall have only positive energy quanta {\small AND} preserve causality {\small AND} be Lorentz\bz{-}covariant. 

In the interaction picture and on free\bz{-}field level, the field operators can be expanded in series of plain wave operators 
\begin{subequations}\label{oskjgjnsdf}\begin{align}
\phi (x)&=\sum _{\boldsymbol{k},r}\Big(\underbrace{\, ^r\! a_{\boldsymbol{k}}\, ^r\! \widetilde{\phi}^{(-)}_{\boldsymbol{k}}\, e^{-ikx}+\, ^r\! b^\dagger _{\boldsymbol{k}}\, ^r\! \widetilde{\phi}^{(+)}_{\boldsymbol{k}}\, e^{+ikx}}_{\displaystyle\, ^r\! \phi _{\boldsymbol{k}}(x)}\Big)\displaybreak[1]\\
\pi (y)&=\sum _{\boldsymbol{f},s}\Big(\underbrace{\, ^sb_{\boldsymbol{f}}\, ^s\widetilde{\pi}^{(-)}_{\boldsymbol{f}}\, e^{-ify}+\, ^s\! a^\dagger _{\boldsymbol{f}}\, ^s\widetilde{\pi}^{(+)}_{\boldsymbol{f}}\, e^{+ify}}_{\displaystyle\, ^s\!\pi _{\boldsymbol{f}}(y)}\Big)\ ,  
\end{align}\end{subequations} 
in which $r,s$ are polarization indices, and $\boldsymbol{k},\boldsymbol{f}$ are the infinitely many wavenumbers which are compatible with the boundary conditions of a finite normalization volume. The quantization rule \eqref{ksmgnsfngf} translates into this 
\begin{align}
&\text{algebra of boson (fermion) Fourier-operators:}\notag\\ 
&\, ^s\! a_{\boldsymbol{f}}\, ^r\! a^\dagger _{\boldsymbol{k}}\mklammerp\, ^r\! a^\dagger _{\boldsymbol{k}}\, ^s\! a_{\boldsymbol{f}}=\, ^s\! b_{\boldsymbol{f}}\, ^r\! b^\dagger _{\boldsymbol{k}}\mklammerp\, ^r\! b^\dagger _{\boldsymbol{k}}\, ^s\! b_{\boldsymbol{f}}=\delta _{\boldsymbol{kf}}\,\delta _{rs} \label{owhsdgbsdfg}\\ 
&\text{All other (anti)commutators are zero.}\notag 
\end{align} 
If an infinite normalization volume is used, then the sums over wavenumbers are replaced by integrals, and the Kronecker\bz{-}symbol of wave\bz{-}numbers is replaced by a delta\bz{-}function. 

For each wavenumber $\boldsymbol{k}$ and each polarization $r$ there is one particle oscillator with creation operator $\, ^r\! a^\dagger _{\boldsymbol{k}}$ and annihilation operator $\, ^r\! a_{\boldsymbol{k}}\, $, and one antiparticle oscillator with creation operator $\, ^r\! b^\dagger _{\boldsymbol{k}}$ and annihilation operator $\, ^r\! b_{\boldsymbol{k}}\, $. With the notation $|n_{a\boldsymbol{k}r}m_{b\boldsymbol{f}s}\rangle $ for a state, in which $n_a$ particles with wavenumber $\boldsymbol{k}$ and polarization $r$, and $m_b$ antiparticles with wavenumber $\boldsymbol{f}$ and polarization $s$ are excited, this additional rule must be stipulated to bound the sequence of oscillator levels from below: 
\begin{align}
\, ^r\! a_{\boldsymbol{k}}|0_{a\boldsymbol{k}r}m_{b\boldsymbol{f}s}\rangle =
\, ^s\! b_{\boldsymbol{f}}|n_{a\boldsymbol{k}r}0_{b\boldsymbol{f}s}\rangle =0\label{msngsnfsda}
\end{align} 
For the state which is annihilated by \emph{all} annihilation operators, we use the notation $|0\rangle $. Note that \eqref{msngsnfsda} is not a mathematical consequence of the quantization rule \eqref{ksmgnsfngf}, nor a consequence of any other precedent theorem. Instead both \eqref{ksmgnsfngf} and \eqref{msngsnfsda} are indispensable parts of a complete rule of field quantization. The canonical quantization rule \eqref{ksmgnsfngf}\,\&\,\eqref{msngsnfsda} was found by guessing. It is a law of nature, designed to match the experimentally observed fact that the energies of all fields are quantized, and that the energy quanta of all fields are positive (no field can lower it's energy due to excitation of quanta). The law \mbox{\eqref{ksmgnsfngf}\,\&\,\eqref{msngsnfsda}} does meet these objectives, but it doesn't match observed reality with respect to the vacuum expectation values of the energydensity\bz{-}stress\bz{-}tensors of elementary fields. 

We will use the abbreviation ES\bz{-}tensor for the energydensity\bz{-}stress\bz{-}tensor. It's components are  
\begin{align}
\mathcal{T}_{\mu\nu}\equiv\sum _z\frac{\partial\mathcal{L}}{\partial (\partial ^\mu\phi ^{(z)}_\rho )}\,\partial _\nu\phi ^{(z)}_\rho -g_{\mu\nu}\mathcal{L}\ .\label{ksmgnmsdfg} 
\end{align} 
The summation $\sum _{\, z}$ is over all fields and anti\bz{-}fields contained within $\mathcal{L}$.  $(g_{\mu\nu})$ is the metric tensor. Greek characters $\mu ,\nu ,\rho ,\dots $ are space\bz{-}time indices. They have to be summed over $0,1,2,3$ automatically, whenever they show up twice in a product. For space\bz{-}like indices latin characters $j,k,l,\dots $ are used, which have to be summed over $1,2,3$ automatically whenever they show up twice in a product. Latin characters $a,b,c,\dots $ are used to specify components which may be spinor components or space\bz{-}time components. 

If boson (fermion) fields are quantized according to \mbox{\eqref{ksmgnsfngf}\,\&\,\eqref{msngsnfsda}}, then the volume integrals and the zero\bz{-}point expectation values of their ES\bz{-}tensor components become\footnote{See \ggeg  sections 15.3, 15.7, 16.2, 17.2 in \cite{Gruendler:fieldtheory}.} in linear approximation\footnote{With exception of the Higgs field, all fields of the standard model of elementary particles are linear fields. Hence \eqref{mfjhfnbfbgfdg} is exact for these fields.} 
\begin{subequations}\label{mfjhfnbfbgfdg}\begin{align}
&\inte _\Omega\!\dif ^3\! x\;\mathcal{T}_{\mu\nu}=F_2\sum _{\boldsymbol{k},r}\frac{c\hbar k_\mu k_\nu }{\sqrt{k_jk^j+m^2c^2/\hbar ^2}}\,\cdot\notag\\ 
&\hspace{4em}\cdot\Big(\, ^r\! a^\dagger _{\boldsymbol{k}}\, ^r\! a_{\boldsymbol{k}}+\, ^r\! b^\dagger _{\boldsymbol{k}}\, ^r\! b_{\boldsymbol{k}} \pklammerm (\underbrace{\, ^r\! b_{\boldsymbol{k}}\, ^r\! b^\dagger _{\boldsymbol{k}}\mklammerp\, ^r\! b^\dagger _{\boldsymbol{k}}\, ^r\! b_{\boldsymbol{k}}}_{1})\Big)\label{odjgjnsdgss}\displaybreak[1]\\    
&\langle 0|\,\mathcal{T}_{\mu\nu}|0\rangle =\frac{F}{\Omega}\sum _{\boldsymbol{k}}\frac{c\hbar k_\mu k_\nu }{\sqrt{k_jk^j+m^2c^2/\hbar ^2}}\label{pogfkgkjmgfj}\displaybreak[1]\\ 
&F\equiv F_1\cdot F_2\cdot F_3\notag\\ 
&F_1=+1\text{ for bosons}\ ,\ F_1=-1\text{ for fermions}\notag\\ 
&F_2=1/2\text{ if field and antifield are identical}\ ,\ F_2=1\text{ else}\notag\\ 
&F_3=\text{number of the field's polarization degrees of freedom\ .}\notag 
\end{align}\end{subequations} 
$\sum _{\,\boldsymbol{k}}$ is the sum over all wavenumbers $\boldsymbol{k}$, which are compatible with the boundary conditions of the finite normalization volume $\Omega $. If an infinite normalization volume is chosen, then the sums over $\boldsymbol{k}$ are replaced by integrals. In case of elementary fields, \eqref{pogfkgkjmgfj} will diverge, no matter whether the normalization volume is chosen finite or infinite. 

Serious problems emerge, if \eqref{pogfkgkjmgfj} is combined with the field equation\!\cite{Einstein:KosmolKonst} 
\begin{align}
R_{\mu\nu}(x)-\frac{R(x)}{2}\, g_{\mu\nu}(x)+\Lambda\, g_{\mu\nu}(x)=-\frac{8\pi G}{c^4}\,\mathcal{T}_{\mu\nu}(x) \label{lkdynjbgf}
\end{align}
of General Relativity Theory (GRT). The Ricci\bz{-}tensor $(R_{\mu\nu})$ and it's contraction $R$ are representing the curvature of space\bz{-}time. $\Lambda $ is the cosmological constant, $(g_{\mu\nu})$ is the metric tensor, and $(\mathcal{T}_{\mu\nu})$ is the ES\bz{-}tensor of all fields contained within space\bz{-}time, \ggie  of all fields existing at space\bz{-}time point $x$ with exception of the metric field $(g_{\mu\nu})$. 

According to astronomical observations\!\cite{Bennet:WMAP9y,Planck_coll:cosmres2015}, the universe can on large scales be well described by the \mbox{$\Lambda $CDM} model\!\cite{Bartelmann:cosmology} with dark energy parameter $\Omega _\Lambda =0.69$ and Hubble parameter $H=68\,\text{km}/(\text{s}\,\text{Mpc})$, resulting into the small positive vacuum energy density 
\begin{align} 
\mathcal{T}_{00}^{\,\text{vacuum}}=\frac{3H^2\Omega _\Lambda c^2}{8\pi G}=5.4\cdot 10^{-10} \,\mathrm{J/m^3}\ .\label{eq:energ2dichte}
\end{align}  
The observational data are furthermore indicating that the universe is in the vast empty regions far\bz{-}off mass concentrations almost flat, possibly a perfectly flat euclidean space. Therefore in this article the untypical isolated spots nearby mass concentrations with significant curvature of space will be ignored. Instead only the typical areas of intergalactic vacuum will be considered, which are described by the metric 
\begin{align}
(g_{\mu\nu})\hspace{-.4em}\ggstackrel[1.3]{\stackrel{\scriptstyle\text{intergalactic}}{\text{vacuum}}}\hspace{-.4em}\text{diagonal}(+1,-a^2,-a^2,-a^2)\\ 
a(t)&=\text{cosmic scale factor}\ .\notag   
\end{align} 
Note that we choose triple\bz{-}minus convention for the metric. With the usual choice $a(\text{today})\equiv 1$, and  \begin{align}
\frac{\dif a}{\dif t}\approx aH\approx 2\cdot 10^{-18}s^{-1} 
\end{align} 
being negligible on laboratory timescales, in excellent approximation Minkowski metric can be applied:  
\begin{align}
(g_{\mu\nu})\hspace{-.6em}\stackrel{\stackrel{\scriptstyle\text{intergalactic}}{\text{vacuum}}}{\approx}\hspace{-.6em}(\eta _{\mu\nu})=\text{diagonal}(+1,-1,-1,-1) 
\end{align} 

It's a plausible physical assumption that the vacuum is isotropic, and that consequently all off\bz{-}diagonal elements of $(\langle 0|\,\mathcal{T}_{\mu\nu}|0\rangle )$ vanish. Let $\text{P}\equiv\langle 0|\,\mathcal{T}_{11}|0\rangle =\langle 0|\,\mathcal{T}_{22}|0\rangle =\langle 0|\,\mathcal{T}_{33}|0\rangle $ be the isotropic vacuum pressure of an arbitrary elementary field. Now consider $(-\eta _{\mu\nu}\text{P})$. As $(\eta _{\mu\nu})$ is a Lo\-rentz\bz{-}covariant tensor and $-\text{P}$ is a Lo\-rentz\bz{-}invariant constant, $(\langle 0|\,\mathcal{T}_{\mu\nu}|0\rangle )$ impossibly could be a Lo\-rentz\bz{-}covariant tensor if it was identical with $(-\eta _{\mu\nu}\text{P})$ in all components with exception of $\langle 0|\,\mathcal{T}_{00}|0\rangle \neq -\eta _{00}\text{P}$. Thus combination of the requirement of Lorentz\bz{-}covariance and the assumption of isotropy of the vacuum implies 
\begin{align}
\langle 0|\,\mathcal{T}_{00}|0\rangle =-\langle 0|\,\mathcal{T}_{11}|0\rangle =-\langle 0|\,\mathcal{T}_{22}|0\rangle =-\langle 0|\,\mathcal{T}_{33}|0\rangle \ ,\label{iagnsaygrb} 
\end{align} 
as pointed out by Zeldovich\!\cite{Zeldovich:ccthelpart}. This condition is not at all trivial, because $\langle 0|\,\mathcal{T}_{\mu\nu}\, |0\rangle /F=\eqref{pogfkgkjmgfj}/F$ can never become negative in any diagonal element. Zeldovich\!\cite{Zeldovich:ccthelpart} remarked that cut-off regularization, \ggie  replacing in case of elementary fields the summation limit $|\boldsymbol{k}|=\infty  $ in \eqref{pogfkgkjmgfj} by $|\boldsymbol{k}|_{\text{max}}=B<\infty $ and then considering $\lim _{B\rightarrow\infty}$, is not appropriate, because this method does not change the signs of the spatial or time\bz{-}like terms, and consequently can not meet the condition \eqref{iagnsaygrb}. But he assumed that a relativistically covariant method of regularization would make \eqref{pogfkgkjmgfj} compatible with \eqref{iagnsaygrb}. 

In appendix\;\hyperlink{ta:appa}{A} Zeldovich's assumption is checked and confirmed. There the results of covariant regularization 
\begin{subequations}\label{jmndnfgns}\begin{align} 
\langle 0|\,\mathcal{T}_{\mu\nu}\, |0\rangle\ggstackrel[.3]{\eqref{ksdjmghnfdx}}-\eta _{\mu\nu}\,\frac{F\hbar cK^{4}}{32\pi ^{2}}\lim _{w\rightarrow\infty }\ln\!\Big(\frac{w ^2}{K^2}\Big)\label{jmndnfgnsa}\\ 
\ggstackrel{\eqref{msngnsdfnhgx}}-\eta _{\mu\nu}\,\frac{F\hbar c}{16\pi ^2}\lim _{\kappa\rightarrow\infty }\kappa ^4\,\ln\!\Big(\frac{\kappa}{K_0}\Big)\label{jmndnfgnsb}\\ 
&\hspace*{-3em}K\equiv mc/\hbar >0\ ,\ K_0\equiv (\text{wavenumber-unit})>0\notag 
\end{align}\end{subequations} 
are derived, which meet the condition \eqref{iagnsaygrb}. \eqref{jmndnfgnsa} is the result of dimensional regularization, first published by Akhmedov\!\cite{Akhmedov:vacener}. \eqref{jmndnfgnsb} is the result of Pauli\bz{-}Villars regularization. It is explicated in appendix\;\hyperlink{ta:appa}{A}, why both regularization results are strictly infinite, even in case $m\rightarrow 0\, $. 

With some plausible assumptions, which are explicated in appendix\;\hyperlink{ta:appb}{B}, a finite range for the zero\bz{-}point energy density of an elementary field can be derived from \eqref{jmndnfgnsb}: 
\begin{align}
\langle 0|\,\mathcal{T}_{00}\, |0\rangle\stackrel{\eqref{mksmgmnsdfg}}{\approx}-F\cdot\Big(\, 5\cdot 10^{\, 48}\dots\, 4\cdot 10^{\, 111}\,\Big)\,\frac{\text{J}}{\text{m}^{3}}\label{kmsmjgnsdffg}
\end{align} 
Deviating assertions by \cite{Akhmedov:vacener,Martin:ccp} are refuted in appendix\;\hyperlink{ta:appa}{A}. Comparison of \eqref{kmsmjgnsdffg} with the vacuum energy density concluded from astronomical observations results into the disturbing ratio 
\begin{align}
\frac{\text{theory}}{\text{observation}}=\frac{\eqref{kmsmjgnsdffg}}{\eqref{eq:energ2dichte}}\approx -F\cdot\Big(\, 10^{\, 58}\dots\, 10^{\, 121}\,\Big)\ .\label{oirfkhdjhs} 
\end{align} 
The discrepancy can easily be absorbed by the cosmological constant. Simply assume 
\begin{align}
\Lambda\stackrel{\eqref{eq:energ2dichte},\eqref{lkdynjbgf}}{=}\frac{8\pi G}{c^4}\Big( 5.4\cdot 10^{-10}\, \mbox{Jm}^{-3}-\sum _z\langle 0|\,\mathcal{T}_{00}^{\, (z)}\, |0\rangle\Big) \label{syfhgjnkgfd} 
\end{align} 
with $\sum _{\, z}$ being the sum over all known and unknown fields, and everything is fine. We just need to believe that $\Lambda $ and $\sum _{\, z}\langle 0|\,\mathcal{T}_{00}^{\, (z)}\, |0\rangle $ really mutually compensate (by chance right now, in the present epoch of cosmic evolution) with the breathtaking accuracy of $58\dots 121$ decimal digits. That's of course hard to believe as long as nobody comes up with a physical explanation. 

The pioneers of quantum field theory were from the outset well aware of this problem. Pauli famously\!\cite[page\,842]{Enz:Nullpkten} estimated (due to a cut\bz{-}off of short wavelengths at the classical electron radius) \al that the universe would not even reach to the moon\arp , if that zero\bz{-}point energy would really exist. And in 1927 Jordan and Pauli commented \cite{Jordan:qedladfrei} with regard to the electromagnetic field (my translation): \al It seems to us [\dots ] that --- in contrast to the eigen\bz{-}oscillations in the crystal grid (where both theoretical and empirical reasons are indicating the existence of a zero\bz{-}point energy) --- no reality can be assigned to that `zero\bz{-}point energy'  $h\nu /2$ per degree of freedom in case of the eigen\bz{-}oscillations of the radiation. As one is dealing with regard to the latter with strictly harmonic oscillators, and as that `zero\bz{-}point radiation' can neither be absorbed nor scattered nor reflected, it seems to elude, including it's energy or mass, any method of detection. Therefore it may be the simplest and most satisfactory conception, that in case of the electromagnetic field that zero\bz{-}point radiation does not exist at all.\ar  

Thus the pioneers clearly saw that there is a basic difference in\bz{-}between phonon fields and the quantized electromagnetic field. Why, then, did it not occur to them that consequently different laws of nature should apply to these different types of fields? They had accepted different quantization rules for boson and fermion fields, to avoid obvious contradictions between theory and observation. Consequently they also should have been (and we should be) ready to accept different rules for the quantization of fields which obviously have, or obviously have not, zero\bz{-}point oscillations. The obvious mismatch between QFT and GRT should have guided them to a better law of nature (as will be proposed in section\;\ref{Absch:imprcanquant}), even if they were in those days not yet aware of the inconsistency described in section\;\ref{Absch:inconsistQFT}\,. 

Zeldovich\!\cite{Zeldovich:ccthelpart} appreciated the zero\bz{-}point energies of elementary fields as a chance to justify a cosmological constant $\Lambda\neq 0\, $. In search for problems (\al unfortunately we have run short of crisis lately\arp ), Weinberg\!\cite{Weinb:KosmConst} identified the same subject as an appropriate topic to worry about. He compiled many convincing arguments, why \eqref{syfhgjnkgfd}, though mathematically perfectly correct, is not at all a physically sound remedy for the mismatch \eqref{oirfkhdjhs}, and dubbed the issue \al the cosmological constant problem\arp . See \cite{Carroll:CosConst,Straumann:cosmconstprobl,Padmanabhan:cosmconst,Nobbenhuis:cosmconst,Miao:darkenergy,Martin:ccp} for reviews which update and enlarge Weinberg's work. 

\enlargethispage{-1\baselineskip}Until mid of the nineteen\bz{-}nineties, it was expected that a solution of the cosmological constant problem would explain why the zero\bz{-}point energies of elementary fields and the cosmological constant actually both are zero. When observations of the redshifts of supernovae \cite{Riess:supernovae,Perlmutter:supernovae} indicated an accelerated cosmic expansion, Weinberg \cite{Weinberg:cosmconst2000} split the problem: He named the quest for an explication of the observed small positive cosmological constant \al the new cosmological constant problem\arp , and the appropriate handling of the diverging ES\bz{-}tensors of elementary quantum fields \al the old cosmological constant problem\arp . In section\;\ref{Absch:imprcanquant} an improved method for the quantization of fields is proposed, which --- as a welcome side\bz{-}effect --- solves the old cosmological constant problem. Only a brief comment on the new cosmological constant problem is added in the \al conclusions\ar  section\;\ref{sec:conclus}\,. 

\section{An inconsistency of quantum field theory}\label{Absch:inconsistQFT}
The inconsistency within quantum field theory is quite simple and obvious. As pointed out above, the combination of the requirement of Lorentz\bz{-}covariance and the assumption of isotropy of the vacuum implies for the ES\bz{-}tensors of elementary fields 
\begin{subequations}\label{kjsdngnsdsg}\begin{align}
&\langle 0|\,\mathcal{T}_{00}^{\text{\,elementary}}|0\rangle\stackrel{\eqref{iagnsaygrb}}{=}-\langle 0|\,\mathcal{T}_{11}^{\text{\,elementary}}|0\rangle\, =\notag\\ 
&\hspace{2em}=-\langle 0|\,\mathcal{T}_{22}^{\text{\,elementary}}|0\rangle =-\langle 0|\,\mathcal{T}_{33}^{\text{\,elementary}}|0\rangle \ .\label{kjsdngnsdsga} 
\end{align} 
On the other hand, the signs of all diagonal elements of the ES\bz{-}tensor are identical for all non\bz{-}elementary fields: 
\begin{align} 
&\text{sign}\Big(\langle 0|\,\mathcal{T}_{00}^{\text{\,non\bz{-}elementary}}\, |0\rangle\Big)\stackrel{\eqref{pogfkgkjmgfj}}{=}\text{sign}\Big(\langle 0|\,\mathcal{T}_{11}^{\text{\,non\bz{-}elementary}}\, |0\rangle\Big) =\notag\\
&=\text{sign}\Big(\langle 0|\,\mathcal{T}_{22}^{\text{\,non\bz{-}elementary}}\, |0\rangle\Big) =\text{sign}\Big(\langle 0|\,\mathcal{T}_{33}^{\text{\,non\bz{-}elementary}}\, |0\rangle\Big)\label{kjsdngnsdsgb} 
\end{align}\end{subequations} 
Note that this difference of signs does not indicate a problem of \eqref{kjsdngnsdsgb} with relativistic covariance. Instead the signs of \eqref{kjsdngnsdsga} are forced by the combination of the requirements of isotropy and covariance, while the requirement of isotropy clearly does not exist for non\bz{-}elementary fields. 

Now consider for example the ES\bz{-}tensor of an elementary real Klein\bz{-}Gordan field, and the ES\bz{-}tensor of the phonon field (\ggie  a non\bz{-}elementary field) of an $N$\bz{-}atom solid. Both are boson fields, and for both the parameter $F\textstackrel{\eqref{mfjhfnbfbgfdg}}(+1)\cdot (1/2)\cdot (1)$ applies. Despite this similarity, \eqref{kjsdngnsdsga} is valid for the elementary Klein\bz{-}Gordan field, while \eqref{kjsdngnsdsgb} is valid for the phonon field, no matter how huge the number $N$ may be, provided that $N$ is finite. But a change of signs somewhere in\bz{-}between \al arbitrary huge\ar  and \al infinite\ar  can not be reasonably justified. The difference of signs between \eqref{kjsdngnsdsga} and \eqref{kjsdngnsdsgb} is an unacceptable inconsistency, indicating a basic flaw in the framework of quantum field theory. 

The inconsistency of signs would be harmless and tolerable, if the zero\bz{-}point expectation values of the ES\bz{-}tensor would be zero in case of elementary fields and\bz{/}or in case of non\bz{-}elementary fields. The zero\bz{-}point energy of phonon fields has been experimentally observed and confirmed beyond doubt\cite{Wilks:Helium}. Hence we must have 
\begin{align}
\langle 0|\,\mathcal{T}_{\mu\nu}|0\rangle =0\quad\text{for elementary fields}\: ,\label{osdkjgfjd}
\end{align} 
to remove the inconsistency between \eqref{kjsdngnsdsga} and \eqref{kjsdngnsdsgb}. This can for example be achieved due to \al normal order\arp\cite{Peskin:QFT} of creation\bz{-} and annihilation\bz{-}operators, or by a modification of the rule of canonical field quantization as proposed below in section\;\ref{Absch:imprcanquant}\,. Other solutions may be possible. In any case, the inconsistency \eqref{kjsdngnsdsg} can only be removed by a substantial intervention into the foundations of quantum field theory. 

The inconsistency \eqref{kjsdngnsdsg} could of course be avoided, if the assumption of relativistic covariance of the vacuum would be skipped. Then \eqref{iagnsaygrb} would not be valid, and \eqref{jmndnfgns} could be replaced by some non\bz{-}covariant method of regularization, like cut\bz{-}off regularization, which would avoid the change of signs. This possible scenario has been considered by Nikoli\'{c}\cite[sec.\,6]{Nikolic:prefFrame}: We could for example model the vacuum as filled with particles, like Dirac did in his hole theory\!\cite{Dirac:hole}. In this non\bz{-}covariant vacuum there would exist a preferred reference frame, in which the mean velocity of the vacuum particles is zero. Nikoli\'{c} noted that this assumption may seem less strange, if we remember that in well\bz{-}established cosmological models there exist locally preferred reference frames, \ggie  the reference frames attached to the co\bz{-}moving observers in the FLRW metric\!\cite{Bartelmann:cosmology}. Obviously this method for the elimination of the inconsistency between \eqref{kjsdngnsdsga} and \eqref{kjsdngnsdsgb} would require an even more radical intervention into quantum field theory than the implementation of the improved method of canonical quantization suggested in the next section.
 
\section{An improved law of canonical quantization}\label{Absch:imprcanquant}
The experimental data leave no doubt that the zero\bz{-}point energies of phonon fields really exist, and are correctly described by \eqref{pogfkgkjmgfj}. Thus one is tempted, in order to get rid of the inconsistency \eqref{kjsdngnsdsg}, to reject the regularization result \eqref{jmndnfgns} and stick to the signs of the not regularized matrix elements. But then relativistic covariance is lost. The only way to keep relativistic covariance {\small AND} avoid the inconsistency of signs \eqref{kjsdngnsdsg}, is to assume that the zero\bz{-}point values of the ES\bz{-}tensor components of elementary fields actually are zero. 

Mathematically, the boson matrix elements $\langle 0|\,\mathcal{T}_{\mu\nu}^{\,\text{Klein-Gordan}}\, |0\rangle $ and $\langle 0|\,\mathcal{T}_{\mu\nu}^{\,\text{phonon}}\, |0\rangle $ differ by nothing than the number of oscillation modes, which is finite for the phonon field, but infinite for the elementary Klein\bz{-}Gordan field. Physically, however, there is a further most important difference: The phonon field has a material substrate, namely the atom grid of a solid or molecule. And the state of the phonon field is at the same time the state of motion of the substrate particles. If $\langle 0|\,\mathcal{T}_{\mu\nu}^{\,\text{phonon}}\, |0\rangle $ would be zero for all $\mu $ and $\nu $, that would mean that the substrate particles would be at rest, thus having well\bz{-}defined positions and momenta at the same time, thus violating Heisenberg's indeterminacy relations. 

Since Einstein introduced special relativity theory\!\cite{Einstein:SpezRelTheor}, we learned that no material substrate (the \al ether\arp ) can be assigned to the electromagnetic field, nor to any other elementary field. Phonons are the wavy motions of the substrate particles, while photons, electrons, and all other quanta of elementary fields, are wavy motions of \emph{nothing}. And the indeterminacy relations will not be violated if \emph{nothing} should come to complete rest.  

A field by definition is \al substrate\bz{-}less\arp , if \emph{nothing} remains if that field is not excited. All elementary fields are substrate\bz{-}less fields. If there are no phonons within a solid, then there still are the atoms constituting the solid. The atoms are the material substrate of the phonon field. Besides phonons, further examples for fields with substrates are magnons, excitons, solitons, plasmons, polaritons, \dots , and many others. 

The quantum field theories, which are based on the quantization rule \eqref{ksmgnsfngf}\,\&\,\eqref{msngsnfsda}, are --- as demonstrated by experience --- excellent descriptions of fields with material substrates. But it was an error to believe that exactly the same quantization rule would give a likewise excellent description of fields with no material substrates. The indeterminacy relations do not enforce zero\bz{-}point oscillations of substrate\bz{-}less fields, and obviously, \ggie  as proved by the flat universe, only fields with material substrates are endowed with zero\bz{-}point oscillations. 

All elementary fields are substrate\bz{-}less, and all substrate\bz{-}less fields are elementary. Thus the notion \al substrate\bz{-}less\ar  may seem (and actually is) redundant. Still this notion will be used frequently in the sequel, because it appropriately emphasizes that feature of elementary fields, which is essential for the modified rule of canonical quantization to be presented now.\footnote{\al Canonical quantization of substrate\bz{-}less fields\ar  was the title of the first version of this article. But as of course nobody understood the significance of that title, I changed it to \al Canonical quantization of elementary fields\arp .} 

Only the combination of \eqref{ksmgnsfngf}, which is dealing with the field amplitudes, and \eqref{msngsnfsda}, which is dealing with the quantum oscillators, completely specifies the operator\bz{-}algebra of a quantum field theory. The improved quantization rule again is consisting of one part which is dealing with the field amplitudes, and one part which is dealing with the quantum oscillators. With the notations as defined in \eqref{oskjgjnsdf}, and with $_a$ and $_b$ being space\bz{-}time or spinor component indices, the field amplitudes part is this equal\bz{-}time\pagebreak[1] 
\begin{subequations}\label{lkskjgnfds}\begin{align}
&\text{canonical quantization rule for boson (fermion) fields:}\notag\\ 
&\, ^s\!\pi _{\boldsymbol{f}a}(t,\boldsymbol{x})\, ^r\! \phi ^b_{\boldsymbol{k}}(t,\boldsymbol{y})\mklammerp\, ^r\! \phi ^b_{\boldsymbol{k}}(t,\boldsymbol{y})\, ^s\!\pi _{\boldsymbol{f}a}(t,\boldsymbol{x})\, =\notag\\  
&\hspace{1em}=\begin{cases}0\quad\text{if }\phi\text{ is a substrate-less field }\text{{\small AND} }\, ^r\! \widetilde{\phi}^{(-)\, b}_{\boldsymbol{k}}\! =\!\, ^r\! \widetilde{\phi}^{(+)\, b}_{\boldsymbol{k}}\! =0\hspace{-1em}\\ 
-i\hbar\delta _{\boldsymbol{kf}}\delta _{rs}\delta _{ab}\,\delta ^{(3)}\! (\boldsymbol{x}-\boldsymbol{y})\quad\text{else}\end{cases}\notag\\ 
&\:\text{All other (anti)commutators are zero.}\label{lkskjgnfdsa}  
\end{align} 
Note that this rule --- different from \eqref{ksmgnsfngf} --- must be specified on a per\bz{-}wavenumber and per\bz{-}polarization basis, to allow for some modes to be excited, while other modes are zero (with no zero\bz{-}point energy and pressure). In case of substrate\bz{-}less fields, \eqref{lkskjgnfdsa} translates into the algebra \eqref{owhsdgbsdfg} of Fourier-operators only if $\, ^r\! \widetilde{\phi}^{(-)\, b}_{\boldsymbol{k}}\neq 0\, $ and $\, ^r\! \widetilde{\phi}^{(+)\, b}_{\boldsymbol{k}}\neq 0\, $. To make sure that the algebra of the Fourier\bz{-}operators is well defined in \emph{all} cases, that part of the improved quantization rule, which is dealing with the quantum oscillators, now is split into three sub\bz{-}parts. First we stipulate this 
\begin{align}
&\text{algebra of boson (fermion) Fourier-operators:}\notag\\ 
&\, ^s\! a_{\boldsymbol{f}}\, ^r\! a^\dagger _{\boldsymbol{k}}\mklammerp\, ^r\! a^\dagger _{\boldsymbol{k}}\, ^s\! a_{\boldsymbol{f}}=\, ^s\! b_{\boldsymbol{f}}\, ^r\! b^\dagger _{\boldsymbol{k}}\mklammerp\, ^r\! b^\dagger _{\boldsymbol{k}}\, ^s\! b_{\boldsymbol{f}}=\delta _{\boldsymbol{kf}}\,\delta _{rs}\notag\\ 
&\text{All other (anti)commutators are zero.}\label{lkskjgnfdsb}
\end{align} 
This is identical to \eqref{owhsdgbsdfg}, but here it is part of the quantization rule (\ggie  a law of nature), while \eqref{owhsdgbsdfg} is merely a mathematical consequence of \eqref{ksmgnsfngf}. The second part of the quantum oscillator rule, which is necessary to bound the sequence of oscillator levels from below, is identical to \eqref{msngsnfsda}: 
\begin{align}
\, ^r\! a_{\boldsymbol{k}}|0_{a\boldsymbol{k}r}m_{b\boldsymbol{k}r}\rangle =
\, ^r\! b_{\boldsymbol{k}}|n_{a\boldsymbol{k}r}0_{b\boldsymbol{k}r}\rangle =0\label{lkskjgnfdsc}  
\end{align} 
In addition we stipulate this third part of the quantum oscillator rule:\pagebreak[1] 
\begin{align}
&\mbox{\parbox{.8\linewidth}{All quantum oscillator levels of substrate\bz{-}less fields must be shifted such that the zero\bz{-}point levels of all oscillation modes are zero.}}\notag\\ 
& ^r\! a_{\boldsymbol{k}}^\dagger\, ^r\! a_{\boldsymbol{k}}+\, ^r\! b_{\boldsymbol{k}}^\dagger\, ^r\! b_{\boldsymbol{k}}\pklammerm Y\ \longrightarrow\  \, ^r\! a_{\boldsymbol{k}}^\dagger\, ^r\! a_{\boldsymbol{k}}+\, ^r\! b_{\boldsymbol{k}}^\dagger\, ^r\! b_{\boldsymbol{k}}\label{lkskjgnfdsd} 
\end{align}\end{subequations} 
While $Y=1$ for almost all substrate\bz{-}less fields, the modulus of $Y\in\mathbbm{R}$ is much larger than $1$ in the particular case of the Higgs field, see section\;\ref{sec:higgs}. 

Due to this rule, the quantum oscillator levels of fields with substrates become in case of 
\begin{subequations}\begin{align}
\text{bosons:}\hspace{1.6em}\tfrac{1}{2}\ ,\ &\tfrac{3}{2}\ ,\ \tfrac{5}{2}\ ,\ \tfrac{7}{2}\ ,\ \dots\\ 
\text{fermions:}\ -\tfrac{1}{2}\ ,\ &\tfrac{1}{2} 
\end{align}
while the quantum oscillator levels of substrate\bz{-}less fields become in case of 
\begin{align}
\text{bosons: }\ &0\ ,\ 1\ ,\ 2\ ,\ 3\ ,\ \dots\\ 
\text{fermions: }\ &0\ ,\ 1\ .  
\end{align}\end{subequations} 
Hence $\langle 0|\,\mathcal{T}_{\mu\nu}\, |0\rangle\neq0$ for fields with substrates, while $\langle 0|\,\mathcal{T}_{\mu\nu}\, |0\rangle =0$ for all substrate\bz{-}less fields. 

One might object that the quantization rule \eqref{lkskjgnfds} is nothing other than conventional canonical quantization plus \al normal order\arp , a measure against annoying zero\bz{-}point values which is well\bz{-}known since decades. There are, however, four essential differences: 
\begin{ggitemize} 
\ggitem{First, the four parts of the improved quantization rule \eqref{lkskjgnfds} are consistent, while normal order (and hence vanishing zero\bz{-}point oscillations) is contradicting \eqref{ksmgnsfngf}.} 
\ggitem{Second, \eqref{lkskjgnfds} is not just an ad\bz{-}hoc measure to get rid of divergences. Instead it is motivated and enforced by the necessity to remove the inconsistency of signs between \eqref{kjsdngnsdsga} and \eqref{kjsdngnsdsgb}.} 
\ggitem{Third, normal order would not remove the huge zero\bz{-}point energy offset of the Higgs field (see next section), but \eqref{lkskjgnfdsd} does. Hence \eqref{lkskjgnfds} does match \emph{all} observations, while a rule like \al apply conventional canonical quantization, and apply in addition normal order in case of elementary fields\ar  would not be consistent with the observed fact, that there is no zero\bz{-}point energy of the Higgs field (as proved by the flat universe).} 
\ggitem{Fourth, \eqref{lkskjgnfds} is --- quite different from the mysterious normal order --- a law of nature, based on clear physical arguments: The indeterminacy relations do not enforce zero\bz{-}point oscillations of substrate\bz{-}less fields, and experience (\ggie  the flat universe) proves, that Nature --- as usual --- was economical and did not waste without need infinite amounts of zero\bz{-}point energy to substrate\bz{-}less fields.} 
\end{ggitemize}\vspace{10ex}   

\section{The Higgs field}\label{sec:higgs} 
The Glashow\bz{-}Salam\bz{-}Weinberg model\footnote{See for example \cite[chap.\,29]{Gruendler:fieldtheory} for an elementary introduction to the GSW\bz{-}model of electroweak interactions and the Higgs mechanism.} of electroweak interactions assumes the existence of a complex weak isospin\bz{-}doublet 
\begin{subequations}\label{jsngnsdgbn}\begin{align} 
&\phi (x)\equiv \begin{pmatrix}{\phi}_{3}(x)+i{\phi}_{4}(x)\\ \phi _0+{\phi}_{1}(x)+i{\phi}_{2}(x)\end{pmatrix}\\ 
&{\phi}_{1}(x),\,{\phi} _{2}(x),\,{\phi}_{3}(x),\,{\phi}_{4}(x),\,\phi _0\in\mathbbm{R}\notag 
\end{align} 
with Lagrangian 
\begin{align}
\mathcal{L}&=c^2\hbar ^2(\mbox{d}_\mu\phi ^\dagger )\mbox{d}^\mu\phi +\frac{\mu ^2c^4}{2}\,\phi ^\dagger\phi -\frac{\mu ^2c^4}{4\phi _0^2}\, (\phi ^\dagger\phi )^2\label{kasbngjsdgfdx}\\ 
&=\sum _{i=1}^4 c^2\hbar ^2(\mbox{d}_\mu{\phi}_i)^\dagger \mbox{d}^\mu{\phi}_i-\mu ^2c^4({\phi}_1)^2\, -\notag\\ 
&\quad -\frac{\mu ^2c^4}{4\phi _0^2}\Big(\sum _{i=1}^4{\phi}^2_i\Big) ^2\! -\frac{\mu ^2c^4}{\phi _0}\sum _{i=1}^4{\phi}_i^2\phi _1-\frac{\mu ^2c^4\phi _0^2}{4}\label{kasbngjsdgfd}
\end{align} 
with $\mu\in\mathbbm{R}\, $. The field's potential energy density, displayed as a solid curve in \eqref{jmsdfnjgsd}, is minimal at the finite field amplitude $\phi _0$. Due to an appropriate gauge transformation, the three Goldstone bosons $\phi _2,\phi _3,\phi _4$ mutate to mass\bz{-}terms of the weak vector bosons, and the Lagrangian of the remaining massive field $\phi _1$ becomes 
\begin{align}
\mathcal{L}=c^2\hbar ^2(\dif _\mu\!\phi _1 )\dif ^\mu\!\phi _1 -\mu ^2c^4{\phi _1}^2 -\frac{\mu ^2c^4}{\phi _0}{\phi _1}^3-\frac{\mu ^2c^4}{4\phi _0^2}{\phi _1}^4-\underbrace{\frac{\mu ^2c^4\phi _0^2}{4}}_{V_0}\ .\label{jsngnsdgbnc} 
\end{align}\end{subequations} 
$\phi _1 (x)$ is the excitation of the field $\phi =\eqref{jsngnsdgbn}$ around it's amplitude $\phi _0$ of lowest energy. $\phi _1 (x)$ is the Higgs boson, which probably has been directly observed at the LHC \cite{Atlas:Higgs2012,CMS:Higgs2012}.\pagebreak[1]  
\begin{align}
\raisebox{-12mm}[16mm][16mm]{\begin{overpic}{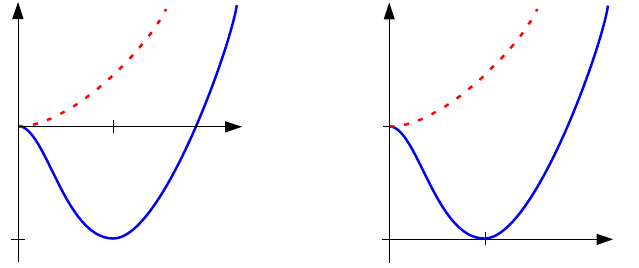}
\put(-3,37){$\scriptstyle V$}
\put(-1.5,20){$\scriptstyle 0$}
\put(-8.5,2.5){$\scriptstyle -V_0$}
\put(37,16.5){$\phi $}
\put(16,16.5){$\phi _{\scriptstyle 0}$}
\put(12,38){\textcolor{red}{$\scriptstyle T>T_c$}}
\put(36.5,31){\textcolor{blue}{$\scriptstyle T<T_c$}}
\put(55,37){$\scriptstyle V$}
\put(54,19.5){$\scriptstyle V_0$}
\put(56.5,3){$\scriptstyle 0$}
\put(95,-0.5){$\phi $}
\put(74,-0.5){$\phi _{\scriptstyle 0}$}
\put(70,38){\textcolor{red}{$\scriptstyle T>T_c$}}
\put(94.5,31){\textcolor{blue}{$\scriptstyle T<T_c$}}
\end{overpic}}\label{jmsdfnjgsd}
\end{align} 
The diagram $V(\phi )$ resembles the diagram $V(M)$ of the free energy density of a ferromagnet with macroscopic magnetization $M$. At high temperature  the macroscopic magnetization is zero. But when the temperature drops below the Curie temperature, the ferromagnet can lower it's free energy due to the formation of a finite macroscopic magnetization. In analogy one is tempted to imagine that there once was a Klein\bz{-}Gordan field with Lagrangian 
\begin{align} 
\mathcal{L}=c^2\hbar ^2(\mbox{d}_\mu \phi ^\dagger )\mbox{d}^\mu \phi -m^2c^4\phi ^\dagger\phi\ ,\label{osksgnsfghs}  
\end{align} 
whose potential energy density is displayed as the dashed curve in \eqref{jmsdfnjgsd}, and that this Lagrangian mutated due to an \al electroweak phase transition\ar  into the Higgs Lagrangian \eqref{kasbngjsdgfdx}, when the temperature of the universe dropped below some critical value $T_c$\,.  

Then one might ask in view of the diagrams \eqref{jmsdfnjgsd}: Should the vacuum not be filled by an easily observable negative energy density $-V_0$ after the electroweak phase transition? Or, if the minimum energy density after the phase transition should be $V=0$ as sketched in the right diagram, doesn't this fact need an explanation? 

Such questions, however, are based on the implicit assumption of a material Higgs ether, while actually the Higgs field is a substrate\bz{-}less field. To understand the argument, read Maxwell's\!\cite{Maxwell:ether} eloquent proof of the reality of the luminiferous ether. Maxwell conceived his theory of the electromagnetic field based on a clear vision of some material substrate, which was to transport electromagnetic waves like a string is transporting mechanical wavy motions. He misunderstood the experimental confirmations of his theory of electrodynamics as evidence for the existence of the ether, and computed the amazing values of that ubiquitous material's density and modulus of elasticity. But the assumption of the ether resulted into unsurmountable problems, and eventually that concept was abandoned. Still Maxwell's equations are a valid and correct description of nature. 

Now compare the Higgs field and the magnetization of a ferromagnet. The constituent particles of the ferromagnetic solid, which have permanent magnetic dipole moments, are the material substrate of the macroscopic magnetization field $M$. The Higgs field, however, has no material substrate, and therefore the analogy with the magnetization must not be taken literally. We must not repeat Maxwell's error and misinterpret the experimental confirmations of the Glashow\bz{-}Salam\bz{-}Weinberg model of electroweak interactions and the direct observation of the Higgs boson as confirmations of some Higgs\bz{-}ether, which would correspond to the ferromagnet. These experimental confirmations prove not more nor less than the existence of a field, which is correctly described by \eqref{jsngnsdgbn}. 

Like the electromagnetic field, the Higgs field exists if and only if minimum one quantum is exited. If it is in the state $|0\rangle $ with no quantum excited, then there exists \emph{nothing}, to which the energy density $V_0$ can be assigned. If the Klein\bz{-}Gordan field \eqref{osksgnsfghs} is in the state $|0\rangle $ of lowest energy ($\phi =0$), then it's energy is zero. If the Higgs field $\phi _1 =\eqref{jsngnsdgbnc}$ is in the state $|0\rangle $ of lowest energy ($\phi =\phi _0$), then it's energy is zero. Thus the Klein\bz{-}Gordan field can not lower it's energy by a phase change into the Higgs field. As a consequence of the law of nature \eqref{lkskjgnfds}, the concept of phase changes is applicable only to material systems, but not to substrate\bz{-}less fields. 

Like the vision of the luminiferous ether helped Maxwell to construct his equations of electrodynamics, the vision of a phase change may have helped the six inventors \cite{Englert:SymBrech,Guralnik:higgs,Higgs:symbrecha,Higgs:symbrechb} of the Higgs field to construct the Lagrangian \eqref{kasbngjsdgfdx}. Still neither the supposed existence of a luminiferous ether nor the supposed historical fact of an electroweak phase change are of any relevance for the validity of Maxwell's equations or the Higgs\bz{-}Lagrangian \eqref{kasbngjsdgfdx}. The only question that matters is, whether or not the consequences drawn from these equations are confirmed by experimental observations. These equations are laws of nature, which stand at the very begin of the respective theories, and don't need colorful justifications. More than a century after Einstein abandoned the ether, we should have learned to trust in the abstract mathematics of well\bz{-}proven laws of nature, and not insist on pictorial explanations.  

As stipulated by the rule \eqref{lkskjgnfdsd}, the large number, which as a consequence of the last term of \eqref{jsngnsdgbnc} is showing up as an adder to $a^\dagger a$ in the Higgs field's ES\bz{-}tensor, must be removed. Note that the ES\bz{-}tensor is the only place where  --- without the rule \eqref{lkskjgnfdsd} --- that term would have any consequences. In the field equation, derived by variation of \eqref{jsngnsdgbnc} with respect to $\phi _1 $, that constant term vanishes anyway. Consequently it would be more elegant, and change nothing of the physics, to replace \eqref{kasbngjsdgfdx} by the equation 
\begin{align}
\mathcal{L}=c^2\hbar ^2(\mbox{d}_\mu\phi ^\dagger )\mbox{d}^\mu\phi +\frac{\mu ^2c^4}{2}\,\phi ^\dagger\phi -\frac{\mu ^2c^4}{4\phi _0^2}\, (\phi ^\dagger\phi )^2+\frac{\mu ^2c^4\phi _0^2}{4}\ ,  
\end{align} 
which is --- like Maxwell's equations --- a law of nature, found by guessing, and not in need for derivation nor explanation. 

While no electroweak phase change is needed for the consistency of the standard model of elementary particles, phase changes of inflaton fields are an indispensable ingredient for many models of cosmic inflations. As explicated above, due to the law of nature \eqref{lkskjgnfds} there are no phase changes of substrate\bz{-}less fields. On the other hand, inflaton fields --- if they exist --- can not have substrates, as proved by the flat universe. Consequently, if one or several cosmic inflations should really have happened (what seems to be very likely, given the astronomical observations), then they either must be caused by something different from phase changes of inflaton fields, or the inflaton fields which caused them must by their very nature be basically different fields, and obey different laws of nature, than the fields of the standard model of elementary particles. 

\section{The perturbative quantum vacuum}\label{Absch:vacfluc} 
While in classical physics a vacuum can be imagined which is a truly empty space, in which no other field than the metric field is existing, the quantum vacuum has a rich structure. For example in quantum electrodynamics, there are in second order of perturbation theory the vacuum bubbles 
\begin{subequations}\label{kfrkjhgnjsg}\begin{align}
\raisebox{-2.4mm}[4mm][3mm]{\includegraphics[trim=5mm 0 6mm 0]{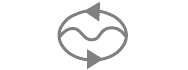}}\quad\text{and}\quad\raisebox{-2.4mm}[4mm][3mm]{\includegraphics{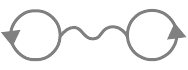}}\quad ,\label{kfrkjhgnjsga}
\end{align}
in fourth order there are bubbles like 
\begin{align}
\raisebox{-2.4mm}[6mm][2mm]{\includegraphics{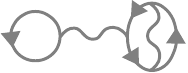}}\quad\text{and}\quad\raisebox{-2.4mm}[4mm][2mm]{\includegraphics{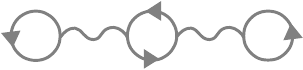}}\quad\text{and}\quad\dots\quad ,
\end{align}\end{subequations}
and so forth. It seems probable, that such bubbles will interact gravitatively. With gravitons represented by wavy double\bz{-}lines, for example structures like this are to be expected: 
\begin{align}
\raisebox{-7mm}{\includegraphics{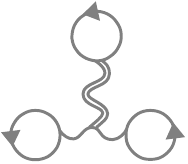}} \label{mksmsgdnsdfg}
\end{align}
Shouldn't such ubiquitous vacuum bubbles induce a significant curvature of space\bz{-}time? 

As we have not yet a full\bz{-}fledged quantum theory of gravitation available by today, this question can not be answered definitively. We can, however, draw useful conclusions from analogies with well\bz{-}known quantum field theories. Consider for example the electromagnetic interaction between two electrically charged fermions. On tree level we get the Feynman diagram 
\begin{subequations}\begin{align}
\raisebox{-4.4mm}{\includegraphics{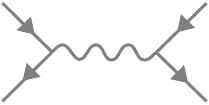}}\quad ,\label{kjdfhgnjdna}
\end{align}
and in fourth order we get diagrams like 
\begin{align}
&\raisebox{-5.4mm}{\includegraphics{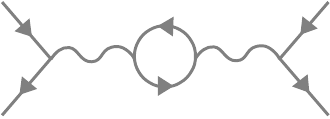}}\quad ,\quad \raisebox{-6.4mm}{\includegraphics{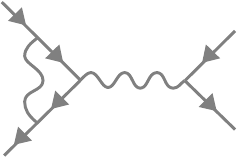}}\quad ,\quad \raisebox{-6.4mm}{\scalebox{-1}[1]{\includegraphics{fourord_2_V3}}}\quad ,\notag\\ 
&\raisebox{-4.4mm}{\includegraphics{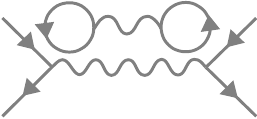}}\quad ,\quad \raisebox{-4.4mm}{\includegraphics{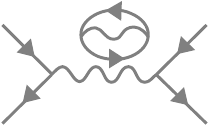}}\quad ,\quad \dots\label{kjdfhgnjdnb} 
\end{align}\end{subequations} 
The essential point is, that the bubbles in the last two diagrams, and the bubbles in all higher orders of perturbation theory, can be factored out and canceled versus the vacuum bubbles \eqref{kfrkjhgnjsg}, see \ggeg  \cite[sec.\,4.4]{Peskin:QFT} or \cite[sec.\,20.1]{Gruendler:fieldtheory}. The same holds true in all other quantum field theories of the standard model of elementary particles. All vacuum bubbles and all disconnected diagrams can be safely ignored in all orders of perturbation theory. The computations lead to correct results, if only connected diagrams like \eqref{kjdfhgnjdna} and the first three diagrams \eqref{kjdfhgnjdnb} are evaluated. 

Note that this is a quite counter\bz{-}intuitive result. Naively one would expect, that vacuum bubbles in the space between the two interacting fermions should add to the dielectric screening. But the theory (which is confirmed experimentally with high precision) tells us, that only connected bubbles like in the first graph of \eqref{kjdfhgnjdnb} add to the net dielectric effect. 

Now consider the gravitative interaction between the same two fermion currents. In the perturbative expansion of this interaction, we will encounter connected diagrams like 
\begin{subequations}\begin{align}
\raisebox{-4.4mm}[5mm]{\includegraphics{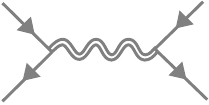}}\quad\text{and}\quad\raisebox{-5.4mm}[5mm]{\includegraphics{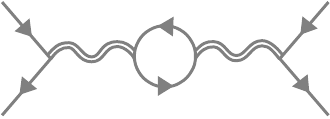}}\quad ,\label{ksmgmndffgha}
\end{align}
and disconnected diagrams like 
\begin{align}
\raisebox{-5mm}[17mm]{\includegraphics{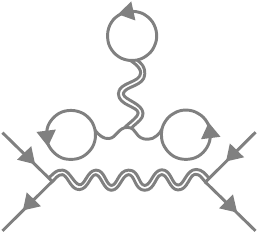}}\quad . 
\end{align}\end{subequations} 
In lack of a complete theory of quantum gravity, it's the most plausible assumption that again all bubbles of all disconnected diagrams can be factored out and canceled versus vacuum bubbles like \eqref{mksmsgdnsdfg}. 

Note that this is a quite counter\bz{-}intuitive result. Naively one would expect, that vacuum bubbles in the space between the two interacting fermions should add to the gravitative attraction. But the theory tells us --- provided that the conclusion by analogy from the known quantum field theories of elementary particles to the yet unknown quantum field theory of gravity is correct --- , that only connected bubbles like in the second graph of \eqref{ksmgmndffgha} add to the net gravitative effect. This implies the most important result, that the curvature of space\bz{-}time is determined exclusively by connected diagrams and by zero\bz{-}point oscillations (if they exist). 

Connected diagrams have minimum one incoming and one outgoing line. Outer lines are representing not virtual particles but real particles on mass\bz{-}shell. Given the very low total energy density of real (observable) particles in the intergalactic vacuum, it's no surprise that they don't induce a measurable curvature of space\bz{-}time. 

Thus the (old) cosmological constant problem is exclusively caused by the zero\bz{-}point oscillations of elementary fields, but not by vacuum bubbles. Once the zero\bz{-}point oscillations have been removed due to the amended law of nature \eqref{lkskjgnfds}, the problem is solved completely, and will not return in higher orders of perturbative treatment. 

\section{And the Casimir effect?}\label{sec:caseff} 
Zero\bz{-}point oscillations have been invoked in computations of many phenomena. The most prominent example is the Casimir effect \cite{Casimir:caseffect}, whose computation is explicated in more detail \ggeg  in \cite{Straumann:cosmconstprobl,Martin:ccp}. For an elementary derivation of Casimir's result, see \cite[sec.\,4]{apin:se08011}. For a review of the theory and experimental proofs of the Casimir effect, see \ggeg \cite{Lambrecht:caseff2}. Many other phenomena can as well be computed in fair approximation due to the assumption of zero\bz{-}point oscillations of substrate\bz{-}less fields, for example the Lamb shift \cite{Martin:ccp}. Is the improved canonical quantization law \eqref{lkskjgnfds} compatible with these phenomena? 

Actually all these phenomena can alternatively be computed by methods, which do not assume zero\bz{-}point oscillations of substrate\bz{-}less fields: Casimir forces between macroscopic surfaces can be computed as van\bz{\:}der\bz{\:}Waals interactions, without invoking zero\bz{-}point oscillations, as demonstrated by Lifshitz\,et.\,al. \cite{Lifshitz:vanderwaal,Dzyaloshinskii:vanderwaal}. And the Lamb shift can be computed\!\cite{Eides:hydrogatoms} by the standard perturbative methods of QED, again without invoking zero\bz{-}point oscillations. 

Thus we are faced with the fact that for all phenomena, which have been computed with reference to zero\bz{-}point oscillations of substrate\bz{-}less fields, there exists an alternative computational method of minimum same prognostic power, which does not resort to zero\bz{-}point oscillations. In a situation where two independent explanations exist for the same phenomena, one of them going without and one of them going with the assumption of zero\bz{-}point oscillations, clearly no stringent conclusion can be drawn regarding the reality of zero\bz{-}point oscillations. Thus, from a logical point of view, we could stop the discussion at this point and simply state, that the experimental confirmation of the Casimir effect is not in conflict with the improved law \eqref{lkskjgnfds} of canonical quantization. Still some further remarks on Casimir's method may be appropriate. 

Jaffe\!\cite{Jaffe:Casimir} named the computations which are based on the assumption of zero\bz{-}point oscillations of elementary fields \al heuristic\arp , thus characterizing this approach as a method which due to artful combination of assumptions, which are not sufficiently substantiated or even wrong, eventually arrives at correct results. The metal plates, which are attracted by the Casimir force, are in Casimir's method represented by boundaries with infinite conductivity. Thus at this point of his computations, Casimir implicitly assumed $\alpha\rightarrow\infty $ for the coupling constant of the electromagnetic field, while in reality the conductivity of the metal plates is finite. But within the same computation he assumed that there is no exchange of virtual photons between the plates, \ggie\ he implicitly assumed $\alpha\rightarrow 0$\,, while in reality virtual photons are coupling to fluctuating currents in the metal plates. Due to artful heuristic combination of these two wrong --- and extremely opposite --- implicit assumptions, Casimir eventually achieved the correct result. 

It's instructive to see how Casimir arrived at his method: In the fall of 1947, Casimir and Polder\cite{Casimir:polder} computed the retarded van\;der\;Waals force between two atoms without permanent dipole moments at large distance $R$. As a preparatory first step, they investigated a simpler setup, in which a single atom with polarizability $\beta $ is placed at a distance $R$ from a metal plane with infinite conductivity. According to classical theory, an interaction between the atom and it's mirror picture was expected, and this was by and large the case. The result was 
\begin{subequations}\begin{align}  
F=-\frac{3\hbar c\beta }{2\pi R^5}\ .\label{sdkdfbngbfa}
\end{align}  
In the next step, Casimir and Polder found 
\begin{align}  
F=-\frac{161\hbar c\beta _1\beta _2}{4\pi R^8}\label{sdkdfbngbfb} 
\end{align}  
for the retarded van\;der\;Waals force between two atoms with polarizabilities $\beta _1$ and $\beta _2$. While the computation of \eqref{sdkdfbngbfb} was very complicated and tedious, Casimir and Polder were surprised by the simplicity, with which they had arrived at the result \eqref{sdkdfbngbfa}. Hence Casimir wondered whether the force could be computed with similar simplicity, if \emph{both} atoms were replaced by metal plates with infinite conductivity, \ggie  by boundaries. But while in case of \eqref{sdkdfbngbfa} the fluctuating dipole moment of the atom had supplied the electromagnetic field which is inducing the van\;der\;Waals interaction, where should the field come from in case of two boundaries? No electromagnetic field is associated with a boundary. Casimir got the essential hint when he was chatting in those days with Bohr about his actual activities. \al Bohr mumbled something about zero\bz{-}point energy\arp , remembered Casimir many years later\cite{Casimir:bohr}. This tip was sufficient for Casimir, to find out that the electromagnetic field's zero\bz{-}point oscillations could supply the missing electromagnetic field, and to compute\!\cite{Casimir:caseffect} the force 
\begin{align}
F_{\text{Casimir}}=-\frac{\pi ^2\hbar c\, A}{240\, R^4}\label{sdkdfbngbfc} 
\end{align}\end{subequations} 
between two perfectly conducting metal plates with area $A$ at distance $R$. 

Nowhere in the derivations of \eqref{sdkdfbngbfa} and \eqref{sdkdfbngbfb} had Casimir and Polder resorted to zero\bz{-}point oscillations and zero\bz{-}point energy of the electromagnetic field. Note furthermore, that the derivation of \eqref{sdkdfbngbfc} is mathematically almost identical to the derivation of \eqref{sdkdfbngbfa}. Thus Casimir actually did nothing essentially different in the derivation of \eqref{sdkdfbngbfc}, but only replaced the fluctuating and\bz{/}or induced dipoles of the atoms by the assumed zero\bz{-}point oscillations of the electromagnetic field. 

(Virtual) photons interact with electrical charges, but zero\bz{-}point oscillations of the electromagnetic field don't. Hence the interaction was lost when Casimir skipped the virtual photons from the picture (\ggie  implicitly assumed $\alpha\rightarrow 0$), and only kept the zero\bz{-}point oscillations. But the interaction was re\bz{-}gained when he in addition changed the metal plates to boundaries (\ggie  implicitly assumed $\alpha\rightarrow\infty $), because the spectrum of zero\bz{-}point oscillations is shaped by the geometry of the boundaries.  

While it is not obvious whether the field in\bz{-}between the plates are virtual photons or zero\bz{-}point oscillations, it's pretty clear and beyond doubt that the plates really are metal plates with finite conductivity, but not boundaries with infinite conductivity. Consequently the picture of virtual photons interacting with fluctuating currents in metal plates does correctly represent the actual situation, while the the picture with zero\bz{-}point oscillations inbetween boundaries is merely an artful substitution, which --- due to the intricate heuristic combination of $\alpha\rightarrow 0$ and $\alpha\rightarrow\infty $ --- leads to the same mathematical result. 

A precise analysis of Casimir's heuristic trick was presented by Nikoli\'c\!\cite{Nikolic:Casimir}. He proved that the wrong impression, that the Casimir force might be somehow related to zero\bz{-}point oscillations, slipped in due to illegitimate manipulations of the equation of motion. 

Thus Casimir's heuristic method of computation can at best be considered an ingenious alternative algorithm for the computation of the van\bz{\,}der\bz{\,}Waals interaction between metal plates. But the success of his method is definitively not indicating the reality of zero\bz{-}point oscillations of substrate\bz{-}less fields. 

\section{Conclusions}\label{sec:conclus} 
The canonical quantization rule \eqref{lkskjgnfds} has been proposed as an improved law of nature, which --- different from the conventional law \mbox{\eqref{ksmgnsfngf}\,\&\,\eqref{msngsnfsda}} --- removes the inconsistency of QFT pointed out in equation \eqref{kjsdngnsdsg}. The essential point is, that there are no zero\bz{-}point oscillations and no zero\bz{-}point energies of elementary (\ggie  substrate\bz{-}less) fields. We emphasized that this assertion is not in conflict with the experimentally confirmed Casimir force, which actually is a van\;der\;Waals force, induced by fluctuating electromagnetic dipoles and transmitted by virtual photons. 

$\langle 0|\,\mathit{T}_{\mu\nu}\, |0\rangle =0$ for all elementary fields, because they are substrate\bz{-}less fields. It has been shown in particular that this as well is true for the Higgs field. For classical fields, the state of lowest energy is the state with zero energy. Non\bz{-}elementary quantum fields can not assume the state with zero energy, because they have substrates, and Heisenbergs indeterminacy relations would be violated if the substrate particles would come to complete rest. No such restriction exists for elementary quantum fields, because they are substrate\bz{-}less fields. We should consider $\langle 0|\,\mathit{T}_{\mu\nu}^\text{elementary}|0\rangle =0$ as the regular standard case, while $\langle 0|\,\mathit{T}_{\mu\nu}^\text{non-elementary}|0\rangle\neq 0$ is the exception, enforced by the fact that the non\bz{-}elementary field's substrate is subject to Heisenbergs indeterminacy relations. 

Plausible arguments have been presented for the assertion, that disconnected perturbative structures of quantum fields, \ggie  vacuum bubbles, have no net gravitative effect, and consequently elementary fields don't contribute to the energy density of the vacuum on any level of perturbative treatment. Thus the improved quantization rule \eqref{lkskjgnfds} does not only remove the inconsistency of QFT, but also solves the old cosmological constant problem. 

The new cosmological constant problem remains. I don't think, however, that the word \al problem\ar  is appropriate. The old cosmological constant problem was --- as displayed in \eqref{oirfkhdjhs} --- an inconsistency of $58\dots 121$ decimal digits within the framework of well\bz{-}established physical theories, and thus indeed a disturbing problem. But the question: \al How can the value 
\begin{align}
\Lambda\stackrel{\eqref{syfhgjnkgfd}}{=}\frac{8\pi G}{c^4}\, 5.4\cdot 10^{-10}\, \mbox{Jm}^{-3}=1.1\cdot 10^{-52}\,\text{m}^{-2}
\end{align} 
of the cosmological constant be explained?\ar  is not more nor less a \al problem\ar  than questions like: Why does the electron have the electrical charge $e=-1.6\cdot 10^{-19}\text{C}$, why is the reduced constant of action $\hbar =1.05\cdot 10^{-34}\text{Js}$, why is $G=6.67\cdot 10^{-11}\text{m}^3/(\text{kg\,s}^2)$ the value of the constant of gravitation? These questions do not indicate inconsistencies of any theory, but merely open topics and objectives for future scientific work. In the basic theories of physics there are more than two dozen constants, whose values we can not yet explain. The anthropic principle \cite{Carter:anthrprinc} \al these constants have the values they have, because we wouldn't be here and ask, if they had significantly different values\ar  may for a long time stay the only answer known to us, unless substantial pieces of \al new physics\ar  will be detected.\vspace{4ex}    

\section*{Appendix A: Regularization \texorpdfstring{of $\boldsymbol{\langle 0|\,\mathcal{T}_{\mu\nu}\, |0\rangle}$}{}}
To \raisebox{4\baselineskip}[0pt][0pt]{\hypertarget{ta:appa}{}}regularize \eqref{pogfkgkjmgfj}, it's advantageous to switch to an infinite normalization volume, and apply spherical coordinates: 
\begin{align}
\langle 0|\,\mathcal{T}_{\mu\nu}\, |0\rangle\ggstackrel[-.15]{\eqref{pogfkgkjmgfj}}F\hbar c\!\inte _0^{+\infty }\hspace{-.4em}\frac{\dif\! k}{(2\pi )^3}\; 4\pi k^2\,\cdot\notag\\ 
&\quad\cdot\bigg( \eta _{\mu 0}\eta ^0{}_{\nu}\,\frac{k^2+K^2}{\sqrt{k^2+K^2}}-\frac{\eta _{\mu j}\eta ^j{}_{\nu}\, k^2}{3\sqrt{k^2+K^2}}\bigg)\notag\\ 
k&\equiv\sqrt{k_jk^j}\quad ,\quad K^2\equiv m^2c^2/\hbar ^2\label{sdmfmdfmhnmn}
\end{align} 
Dimensional regularization\footnote{See \cite[sections\,25.4\,and\,30.3]{Gruendler:fieldtheory} for computational details.} with $D=3-\epsilon $ and $0<\epsilon\in\mathbbm{R}$: 
\begin{align}
\langle 0|\,\mathcal{T}_{\mu\nu}\, |0\rangle &=\lim _{D\rightarrow 3}\frac{F\hbar c}{(2\pi )^D}\, \frac{2\pi ^{D/2}}{\Gamma (D/2)}\!\inte _0^{+\infty }\!\dif\! k\, k^{D-1}\,\cdot\notag\\ 
&\quad\cdot\bigg[ \frac{\eta _{\mu 0}\eta ^0{}_{\nu}3(k^2+K^2)-\eta _{\mu j}\eta ^j{}_{\nu}k^2}{3\sqrt{k^2+K^2}}\bigg] 
\end{align} 
If 
\begin{subequations}\label{kdfjgjhxgf}\begin{align}
K\equiv mc /\hbar >0\ ,  
\end{align} 
then the substitution   
\begin{align} 
k=K\!\raisebox{-2.8mm}{\,\scalebox{.6}{$+$}\!}\sqrt{\frac{1-X}{X}}\quad\text{with}\ X\equiv\frac{K^2}{k^2+K^2} 
\end{align}\end{subequations} 
is possible, resulting into  
\begin{align}
&\langle 0|\,\mathcal{T}_{\mu\nu}\, |0\rangle =\lim _{\epsilon\rightarrow 0^+}\frac{F\hbar cK^{4}}{24\pi ^{3/2}\Gamma (3/2)}\cdot\Big(\frac{4\pi}{K^2}\Big) ^{\epsilon /2}\Gamma (-2+\epsilon /2)\,\cdot\notag\\ 
&\quad\cdot\bigg[\frac{\eta _{\mu 0}\eta ^0{}_{\nu}3\Gamma (3/2)}{\Gamma (-1/2)}- 
\frac{\eta _{\mu j}\eta ^j{}_{\nu}\Gamma (5/2)}{\Gamma (+1/2)}\bigg]\ .   
\end{align} 
With the relations 
\begin{subequations}\begin{align} 
&\Gamma (\epsilon /2-n)=\Gamma (\epsilon /2)\cdot\prod _{\nu =1}^n\frac{1}{\epsilon /2-\nu}\quad ,\quad n=1,2,3,\dots\label{oadsgnhsgfrb}\displaybreak[1]\\ 
&\Gamma (\epsilon /2)=\frac{2}{\epsilon}-\gamma +\mathcal{O}(\epsilon ^2)\quad ,\quad\gamma\equiv 0.577215\dots\displaybreak[1]\\ 
&\Big(\frac{4\pi\, K_0^2}{K^2\, K_0^2}\Big) ^{\epsilon /2}\hspace{-.4em}=(K_0^2)^{-\epsilon /2}\Big[ 1+\frac{\epsilon}{2}\,\ln\Big(\frac{4\pi K_0^2}{K^2}\Big)\, +\mathcal{O}(\epsilon ^2)\Big] 
\end{align}\end{subequations}
one arrives at 
\begin{subequations}\label{ksdjmghnfdxyy}\begin{align} 
\langle 0|\,\mathcal{T}_{\mu\nu}\, |0\rangle &=-\eta _{\mu\nu}\,\frac{F\hbar cK^{4}}{32\pi ^{2}}\,\lim _{\epsilon\rightarrow 0^+}\Big[\frac{2}{\epsilon}-\gamma +\ln\Big(\frac{4\pi K_0^2}{K^2}\Big)\Big]\label{ksdjmghnfdxyya}\\ 
&=-\eta _{\mu\nu}\,\frac{F\hbar cK^{4}}{32\pi ^{2}}\lim _{w\rightarrow\infty }\ln\!\Big(\frac{w ^2}{K^2}\Big)\label{ksdjmghnfdx}\\ 
K&=mc/\hbar >0\ .\notag 
\end{align}\end{subequations} 
In the last step, modified minimal subtraction $\overline{\text{MS}}$ was applied. This result was first published by Akhmedov \cite{Akhmedov:vacener}.

Because of the substitution \eqref{kdfjgjhxgf}, the result \eqref{ksdjmghnfdxyy} is not valid for $K=mc/\hbar =0$. We can however consider the limit of arbitrary small --- and even unmeasurable small --- mass. Different from the assertions of \cite{Akhmedov:vacener,Martin:ccp}, the result \eqref{ksdjmghnfdxyy} stays infinite also in this case, for the following reason. One can find physical arguments to stop the limit ${B\rightarrow\infty }$ at some finite cut\bz{-}off\bz{-}wavenumber $B_{\text{max}}$ in cut\bz{-}off regularization, or to stop the limit ${\kappa\rightarrow\infty }$ at some finite invariant wavenumber $\kappa _{\text{max}}$ of the Pauli\bz{-}Villars counterterm (see appendix\;\hyperlink{ta:appb}{B}). But there is no physical argument known to stop $\epsilon\rightarrow 0$ (and consequently $w\rightarrow\infty  $) in dimensional regularization at any finite value $\epsilon _{\text{min}}$. Instead $\epsilon $ must become strictly zero, and thus the result \eqref{ksdjmghnfdxyy} is strictly infinite, even if a field's mass $m$ and invariant wavenumber $K$ should be unmeasurable small. 

For the same reason it would not be sensible to say --- as done by \cite{Akhmedov:vacener,Martin:ccp} --- that the logarithmic divergence of \eqref{ksdjmghnfdx} is less severe than the fourth\bz{-}power divergence encountered with cut\bz{-}off regularization. One can reasonably state $B_{\text{max}}^4/K_0^4\gg\ln (B_{\text{max}}/K_0)$, but it's not reasonable to discuss $\infty \,\textstackrel[\gg ]{\boldsymbol{?}}\,\infty  $. Furthermore remember that the divergence in dimensional regularization is \emph{always} logarithmic because of \eqref{oadsgnhsgfrb}, no matter which horrible powers of $k$ are showing up in a diverging integral.  

Pauli-Villars regularization\footnote{See \cite[sections\,25.3\,and\,30.3]{Gruendler:fieldtheory} for computational details.}: First the ES\bz{-}tensor's spatial components are regularized. The insertion of one counter\bz{-}term is sufficient.\pagebreak[1]  
\begin{align}
\langle 0|\, &\mathcal{T}_{il}\, |0\rangle\stackrel{\eqref{sdmfmdfmhnmn}}{=}\notag\displaybreak[1]\\ 
&=-\eta _{ij}\eta ^j{}_{l}\frac{F\hbar c}{6\pi ^2}\!\inte _0^{+\infty }\!\dif k\,\bigg( \frac{k^4}{\sqrt{k^2+K^2}}-\lim _{\kappa\rightarrow\infty }\frac{k^4}{\sqrt{k^2+\kappa ^2}}\bigg)\, =\notag\displaybreak[1]\\ 
&=-\eta _{ij}\eta ^j{}_{l}\frac{F\hbar c}{16\pi ^2}\bigg(\! -K^4\,\ln\!\Big(\frac{K}{K_0}\Big) +\lim _{\kappa\rightarrow\infty }\kappa ^4\,\ln\!\Big(\frac{\kappa}{K_0}\Big)\bigg)\label{msngnsdfnhg}\\ 
K&\equiv mc/\hbar >0\ ,\ K_0\equiv (\text{wavenumber-unit})>0\notag
\end{align} 

As the Pauli\bz{-}Villars method is applicable only to integrals with negative powers of the invariant wavenumber $K$, \eqref{msngnsdfnhg} must be constrained to $m>0\, $, and the ES\bz{-}tensor's $00$\bz{-}component can not at all be regularized by that method. But as \eqref{msngnsdfnhg} is a correct result, it can be combined with the condition \eqref{iagnsaygrb}, and thus this general result can be concluded:  
\begin{align}
\langle 0|\,\mathcal{T}_{\mu\nu}\, |0\rangle\ggstackrel[.58]{\eqref{sdmfmdfmhnmn},\eqref{iagnsaygrb}}-\eta _{\mu\nu}\,\frac{F\hbar c}{16\pi ^2}\bigg(\lim _{\kappa\rightarrow\infty }\kappa ^4\,\ln\!\Big(\frac{\kappa}{K_0}\Big) -K^4\,\ln\!\Big(\frac{K}{K_0}\Big) \bigg)\notag\\ 
&=-\eta _{\mu\nu}\,\frac{F\hbar c}{16\pi ^2}\lim _{\kappa\rightarrow\infty }\kappa ^4\,\ln\!\Big(\frac{\kappa}{K_0}\Big)\label{msngnsdfnhgx} \\ 
K&\equiv mc/\hbar >0\ ,\ K_0\equiv (\text{wavenumber-unit})>0\notag 
\end{align} 
Note that the finite term could be neglected. The result is consistent with the result \eqref{ksdjmghnfdxyy} of dimensional regularization. Neither the result of dimensional regularization nor the result of Pauli\bz{-}Villars regularization show any difference for finite $m$ versus $m\rightarrow 0\, $. As both regularization results are strictly infinite, any further comparison would be pointless. 

\section*{Appendix B: Finite values\texorpdfstring{ for $\boldsymbol{\langle 0|\,\mathcal{T}_{\mu\nu}\, |0\rangle}$}{}}
To \raisebox{4\baselineskip}[0pt][0pt]{\hypertarget{ta:appb}{}}extract finite values from the regularized result \eqref{msngnsdfnhgx}, two types of arguments can be applied, which lead to the same result: 

We can argue that the quantum field theories, which led to \eqref{msngnsdfnhgx}, are only low\bz{-}energy effective theories, and therefore the limit $\kappa\rightarrow\infty $ must be replaced by some finite wave\bz{-}number $\kappa _{\text{max}}$, like the integration over phonon wavenumbers is cut off at the maximum $|\boldsymbol{k}|$ of the first Brillouin zone. 

Alternatively we may speculate that there exists a supersymmetric fermion partner field for each boson field and a supersymmetric boson partner field for each fermion field. If the partner fields had same mass and same number of polarizations, then the vacuum values of their ES\bz{-}tensors would mutually compensate due to the different signs of $F_1$ in \eqref{mfjhfnbfbgfdg}. Such supersymmetric partner fields have not yet been observed. Still they might exist, but their masses might due to some symmetry breaking be so large by today, that they are beyond reach of experimental observation. Then the vacuum values of the ES\bz{-}tensors of supersymmetric partners would not compensate to zero but to some finite values, which can be estimated by replacing the diverging wave\bz{-}numbers $\kappa $ of the Pauli\bz{-}Villars counterterms in \eqref{msngnsdfnhgx} with the (unknown) large but finite invariant wave\bz{-}numbers $\kappa _{\text{SSP}}$ of the supposed supersymmetric partner fields. 

The invariant wave\bz{-}numbers $\kappa _{\text{max}}$ or $\kappa _{\text{SSP}}$ hardly can be less than about 
\begin{subequations}\begin{align}
\frac{10^{12}\text{eV}}{\hbar c}=5\cdot 10^{18}\text{m}^{-1}\ , 
\end{align} 
because otherwise the supersymmetric partner fields, or some effect of $\kappa _{\text{max}}$, had been observed in high\bz{-}energy collider experiments. 

An upper limit can be estimated due to Heisenberg's indeterminacy relations: A particle cannot be localized with an accuracy better than half it's reduced Compton wavelength, \ggie  better than half it's inverted invariant wave\bz{-}number. This localization must not be less than two times the Schwarzschild radius $r_S\, $, because otherwise the particle would collapse to a black hole: 
\begin{align}
\Delta x&\approx\frac{1}{2\kappa}\geq 2r_S=\frac{4G}{c^2}\cdot\frac{\kappa\hbar}{c}\notag\\ 
\kappa &\leq\sqrt{\frac{c^3}{8G\hbar }}=\frac{1}{l_{\text{Planck}}\sqrt{8}}=2.2\cdot 10^{34}\text{m}^{-1}  
\end{align}\end{subequations} 
Thereby we get for the vacuum expectation values of the ES\bz{-}tensors of each observed elementary field (and possibly it's unobserved supersymmetric partner) the possible range 
\begin{align}
\langle 0|\,\mathcal{T}_{\mu\nu}\, |0\rangle\approx -\eta _{\mu\nu}F\cdot\Big(\, 5\cdot 10^{48}\dots\, 4\cdot 10^{111}\,\Big)\,\frac{\text{J}}{\text{m}^{3}}\ .\label{mksmgmnsdfg} 
\end{align} 
\flushleft{\interlinepenalty=100000\bibliography{../gg}} 
\end{document}